\documentclass[12pt]{article}
\usepackage{amssymb}
\usepackage{amsmath}
\usepackage{graphicx}
\usepackage[pagebackref=false,colorlinks=,linkcolor=blue,citecolor=blue]{hyperref}
\usepackage{bm}
\newcommand{\mathsym}[1]{{}}

\usepackage{graphicx}
\usepackage{rotating}
\usepackage{color}
\usepackage{cancel}
\usepackage{pgfplots}
\usepackage{orcidlink}
\usepackage{doi}
\newcommand{\rmd}{{\rm d}}
\newcommand{\rmi}{{\rm i}}
\newcommand{\rme}{{\rm e}}
\newcommand{\bra}{\begin{array}}
\newcommand{\era}{\end{array}}
\newcommand{\beq}{\begin{equation}}
\newcommand{\eeq}{\end{equation}}
\newcommand{\beqar}{\begin{eqnarray}}
\newcommand{\eeqar}{\end{eqnarray}}

\newcommand{\be}{\begin{equation}}
\newcommand{\ee}{\end{equation}}
\newcommand{\bea}{\begin{eqnarray}}
\newcommand{\eea}{\end{eqnarray}}
\newcommand{\bd}{\begin{displaymath}}
\newcommand{\ed}{\end{displaymath}}
\newcommand{\red }{ \color{black} }

\newcommand{\va }{\varphi}

\newcommand{\lb }{ \left (}
\newcommand{\rb }{ \right )}

\newcommand{\p }{ \partial}

\newcommand{\xh }{ \hat{x}}

\newcommand{\xxh }{ \hat{X}}

\newcommand{\pph }{ \hat{P}}

\newcommand{\ph }{ \hat{p}}

\begin{document}

\title{An algebraic approach to gravitational quantum mechanics}
\author{
Won Sang Chung\footnote{E-mail: mimip44@naver.com}\\
{\small Department of Physics and Research Institute of Natural Science,}\\[-2mm]
{\small College of Natural Science,
 Gyeongsang National University, Jinju 660-701, Korea}\\[5mm]
Georg Junker\footnote{E-mail: georg.junker@fau.de; gjunker@eso.org}\\
{\small Institute f\"ur Theoretische Physik I,
Friedrich-Alexander Universit\"at Erlangen–N\"urnberg,}\\[-2mm]
{\small Staudtstra{\ss}e 7, 91058 Erlangen, Germany;}\\
{\small European Southern Observatory,
Karl-Schwarzschild-Straße 2, 85748 Garching, Germany} \\[5mm]
Hassan Hassanabadi\footnote{E-mail: hha1349@gmail.com}\\
{\small Departamento de F\'{i}sica Te\'{o}rica, At\'{o}mica y Optica and}\\[-2mm]
{\small Laboratory for Disruptive Interdisciplinary Science (LaDIS),}\\[-2mm]
{\small Universidad de Valladolid, 47011 Valladolid, Spain;}\\
{\small Department of Physics, Faculty of Science, University of Hradec Kr\'{a}lov\'{e},}\\[-2mm] {\small Rokitansk\'{e}ho 62, 500 03   Hradec   Kr\'{a}lov\'{e}, Czechia}
}
\date{}
\maketitle

\begin{abstract}
Most approaches towards a quantum theory of gravitation indicate the existence of a minimal length scale of the order of the Planck length.
Quantum mechanical models incorporating such an intrinsic length scale call for a deformation of Heisenberg's algebra resulting in a generalized uncertainty principle and constitute what is called gravitational quantum mechanics.
Utilizing the position representation of this deformed algebra, we study various models of  gravitational quantum mechanics. The free time evolution of a Gaussian wave packet is investigated as well as the spectral properties of a particle bound by an external attractive potential. Here the cases of a box with infinite walls and an attractive potential well of finite depth are considered.
\end{abstract}


\section{Introduction}
In the early 20th century, physics has seen the advent of two fundamental theories, the general relativity being the contemporary theory describing the gravitational force, and the quantum theory being the basis for the description of the electro-weak and strong interactions. Quantum gravity is the attempt to unify both theories into a more fundamental one. Despite the many attempts for such a unification no consistent quantum-gravity theory could be formalised for the time being. In fact, to the best of our knowledge, the first attempt was made by Bronstein in 1936 \cite{Bronstein1936}. Bronstein already concluded that there must be a minimal length in such a theory. Some 10 years later, in 1947, Snyder \cite{1} arrived at the same conclusion by assuming that space-time is not a continuum but discrete. Lorentz invariance then{\red led} him to the conclusion that the usual Heisenberg commutation relation of position operator $\xxh$ and momentum operator $\pph$  calls for a deformation of the form
\be\label{Commutator}
[ \xxh, \pph ] = \rmi ( 1 + \beta \pph^2 )\,.
\ee
Here and throughout this paper we will use units where $\hbar =1$. In the above $\beta > 0 $ represents the deformation parameter which, in our units, has the dimension of an area. Hence, for simplifying notation we will also use the alternative parameter $\lambda=\sqrt{\beta}$ which has the dimension of a length. As quantum gravity effects are believed to become visible at the Planck scale this length should be of the order of the Planck length, $\lambda\sim 10^{-35}\,m$. {\red For a detailed discussion on this parameter we refer to ref.\ \cite{Ref1}.}
Quantum mechanics based on above deformed Heisenberg algebra is called gravitational quantum mechanics.
The deformed algebra implies the uncertainty relation
\be\label{Uncertainty}
  \Delta X \Delta P \ge \frac{1}{2}\left|\langle[ \xxh, \pph]\rangle\right| = \frac{1}{2}\left[ 1+ \lambda^2\langle \pph^2\rangle\right]\,,
\ee
which,{\red see for example \cite{Ref2}}, results in the lower bound
\be\label{Uncertainty2}
  \Delta X  \ge \frac{1}{2}\left[\frac{1}{\Delta P}+ \lambda^2\Delta P\right]\,.
\ee
The minimal length is given by $\min \Delta X = \lambda$ where $\Delta P=1/\lambda$. Thus, to incorporate the concept of a minimum measurable length into quantum mechanics, one should deform the standard Heisenberg algebra in the form \eqref{Commutator}. The resulting uncertainty relation \eqref{Uncertainty} is called Generalized Uncertainty Principle (GUP).

Since the middle of the last century an enormous amount of work has been invested in this topic resulting in a vast number of papers addressing the effects of
GUP on quantum mechanical systems \cite{Ref3,Ref4,Ref5,Ref6}. There had been investigations in the high-energy regime, but also studies of black holes and their thermodynamic properties.
As example we refer to ref.\ \cite{[1]} where the authors analyzed a precise GUP formulation that can be played by the geometry of the momentum space in determining the presence of a minimal length in the theory. {\red Thermodynamic properties of black holes computed via the GUP were also studied in \cite{Ref7}.}
The $n$-dimensional extension of the Kempf et al.\ model \cite{10}, which leads to the appearance of non-commutativity in the configurational variables, is investigated in ref.\ \cite{[2]}.  A comparison between the original GUP and polymer quantum mechanics in isotropic cosmology is performed in \cite{[3]} and in \cite{[4]} the authors studied different versions of the GUP, in regards to the presence of a minimal length and to their implementation to the anisotropic Bianchi I cosmological model. Ref.\ \cite{[5]}, by using an alternative GUP linked to polymer quantum mechanics, obtained a non-singular emergent universe with an asymptotic Einstein-static beginning, and then study corrections to the primordial power spectrum of scalar perturbations during a cosmological-constant-dominated phase. By using this formalism, in ref.\ \cite{[6]}, the gravitational collapse of a spherical dust cloud is studied and finds super-Schwarzschild asymptotic configurations that are also stable to small perturbations. For further previous work, we limit ourselves to refer to some more recent extensive reviews \cite{Hossenfelder2013,Tawfik2015} which provide huge lists of original references to the subject.

The purpose of this paper is to investigate the GUP algebra, that is the deformed Heisenberg algebra \eqref{Commutator} with its generalized uncertainty principle \eqref{Uncertainty}, in an algebraical way. In doing so, we will work within a coordinate representation, where the momentum operator is represented by the so-called GUP derivative. Using the specific properties of this derivative, we are able to study various problems in gravitational quantum mechanics in an explicit way.

We start with reviewing some of the known representations of the GUP algebra in section 2. Section 3 then introduces the aforementioned GUP derivative. We discuss some of its properties and in particular the GUP exponential function being an eigenfunction of this derivative. Continuity conditions for the GUP derivative of wave functions are found in cases where the potential has a finite discontinuity are also suggested.
These properties are then utilized in the following sections. In section 4 we study the free gravitational quantum dynamics of a Gaussian wave packet. Hereby we obtained the GUP Fourier transformation as an approximation of the standard Fourier transformation adapted to the needs of gravitational quantum mechanics, which in turn allows us to discuss in section 5 the time evolution of the initial Gaussian wave package. In section 6 we will look at the eigenvalue problem of the GUP Hamiltonian for the particle in a box with infinite walls and a potential well of finite depth. Some final remarks are provided in section 7.

\section{Representations of the GUP algebra}
The GUP algebra \eqref{Commutator} exhibits various representations, some of which we will discuss below. \\

\noindent {\it GUP momentum representation:}\\
The most natural representation is the one where the operator $\pph$ is represented by a real number $P\in\mathbb{R}$. Consequently, $\xxh$ is then represented by a differential operator. That is, we have
\be\label{PRepresentation}
\xxh = \rmi  \lb 1 + \beta P^2 \rb\frac{\p}{\p P}\,,\qquad \pph=P\,,
\ee
which may easily be verified to obey the GUP algebra \eqref{Commutator}. This representation was first discussed in full detail by Kempf et al \cite{10}. They have shown that the corresponding Hilbert space ${\cal H}_P=L^2\left(\mathbb{R},\rmd\mu(P)\right)$ is equipped with a scalar product of the form
\be\label{inner1}
\langle \Phi| \tilde{\Phi} \rangle =\int_{-\infty}^{\infty} \rmd \mu(P)\,\Phi^* (P)\tilde{\Phi} (P) =\int_{-\infty}^{\infty}  \frac{ \rmd P}{ 1 + \beta P^2  }\Phi^* (P)\tilde{\Phi} (P)\,.
\ee
Here the states are represented by wave functions $\Phi(P)=\langle P|\Phi\rangle$ and $\tilde{\Phi}(P)=\langle P|\tilde{\Phi}\rangle$.
The expectation values of an observable $\hat{A}$ in such state $\Phi(P)$ is given by
\be
\langle \hat{A} \rangle_{\Phi}  = \langle \Phi| \hat{A}|  \Phi\rangle =\int_{-\infty}^{\infty}  \frac{\rmd P}{ 1 + \beta P^2 }\,\Phi^* (P)\bigl(\hat{A} \Phi\bigr) (P)\,.
\ee
Let us also note that both, $\pph$ and $\xxh$, are symmetric operators on ${\cal H}_P$ as was explicitly shown in \cite{10}. The corresponding non-relativistic GUP Hamiltonian of a particle with mass $M>0$ in the presence of a real-valued scalar potential $V$ reads
\be
H_P := \frac{P^2}{2M} + V \lb \rmi  \lb 1 + \beta P^2 \rb\frac{\p}{\p P} \rb \,.
\ee
For an explicit discussion of the eigenvalue problem of $H_P$ for the harmonic oscillator problem we refer the work by Kempf et al \cite{10}.
{\red The harmonic oscillator and the Coulomb problem were also studied by Brown \cite{Ref8}.}
In going forward we will call the representation \eqref{PRepresentation} the GUP momentum representation.\\

\noindent {\it Canonical momentum representation:}\\
Let us consider the standard canonical momentum representation where the momentum operator $\ph$ is represented by a real number $p$ and the position operator $\xh$ by the derivative $i \p_p$.  Obviously we have
\be
\xh = \rmi \p_p\,, \qquad\ph =p\,, \qquad [\xh, \ph]=\rmi\,,
\ee
the standard canonical commutation relation. In terms of this canonical representation the GUP algebra is realized by the relations
\be
\xxh = \rmi \p_p\,,\qquad\pph=\frac{1}{\lambda} \tan \lb \lambda p \rb\,,
\ee
as can easily be verified. We will call this the canonical momentum representation of the GUP algebra.
Whereas the GUP position operator  $\xxh$ and the canonical position operator $\xh$ exhibit the same realization, the corresponding momentum operators are significantly different as $\lambda P= \tan(\lambda p)$. That is, $p$ cannot take arbitrary real values but is restricted to the interval $-p_0<p<p_0$, where
$p_0:= \frac{\pi}{2\lambda}$ represents a cutoff of the canonical momentum. The inverse map is given by the $\arctan$ function definite as the principle branch of the inverse of the tangent function, which we will denote by $\varphi$ wherever suitable,
\be\label{varphi}
p = \varphi(P):= \frac{1}{\lambda} \arctan ( \lambda P)\,.
\ee
Noting that $\varphi'(P)=1 + \beta P^2$ the scalar product \eqref{inner1} for two wave function $\phi(p):=\langle p|\phi\rangle$ and $\tilde{\phi} (p)=\langle p|\tilde{\phi}\rangle$ changes to
\be
\langle \phi| \tilde{\phi}\rangle =\int_{-p_0}^{p_0} \rmd p \,\phi^* (p)\tilde{\phi} (p)\,.
\ee
That is, in essence ${\cal H_P}$ is changed to ${\cal H}_{p}:= L^2([-p_0,p_0],\rmd p)$. Expectation values of an observable $\hat{A}$ in a state represented by a wave function  $\phi(p)$ are given by
\be
\langle \hat{A} \rangle_{\phi}  = \langle \phi| \hat{A}|  \phi\rangle =\int_{-\frac{\pi}{2 \lambda}}^{\frac{\pi}{2 \lambda}}  \rmd p\, \phi^* (p)\bigl(\hat{A} \phi\bigr) (p)
\ee
and the corresponding GUP Hamiltonian reads in this canonical momentum representation
\be
 H_p:= \frac{1}{2M\lambda^2}  \tan^2 \lb \lambda p \rb  + V \lb \rmi  \p_p\rb
\ee

\noindent {\it GUP position representation:}\\
Recalling the well-known relation $[\xh,f(\ph)]=\rmi f'(\ph)$ valid for any reasonable function $f$, the position representation of the GUP algebra is
\be\label{xReps}
\xxh = \xh = x\,, \qquad \pph=\frac{1}{\lambda} \tan \lb \lambda\ph \rb =\frac{1}{\lambda} \tan \lb -\rmi\lambda \p_x \rb = \frac{1}{\rmi\lambda} \tanh \lb \lambda \p_x \rb\,.
\ee
When dealing with GUP quantum mechanics, one usually assumes that $\lambda$ is sufficiently small. Hence, we may express the momentum operator with the help of the Taylor series for the hyperbolic tangent, which reads
\be\label{tanhSeries}
\tanh z =z\sum_{n=0}^{\infty}\frac{2^{2n+2}(2^{2n+2}-1)}{(2n+2)!} \,B_{2n}\,z^{2n}\,,
\ee
where $B_n$ stands for the $n$-th Bernoulli number. Hence, to first order in $\beta=\lambda^2$, the momentum operator is given by
\be\label{Dxapprox}
\pph = -\rmi \p_x + \rmi\frac{\lambda^2}{3} \p_x ^3 +O(\lambda^4)\,.
\ee
In this representation the inner product for two states characterised by the wave functions $\psi(x)=\langle x|\psi\rangle$ and $\tilde{\psi} (x) = \langle x|\tilde{\psi}\rangle$ is the standard product
\be
\langle \psi| \tilde{\psi} \rangle =\int_{-\infty}^{\infty}  \rmd x\, \psi^* (x)\tilde{\psi} (x)\,.
\ee
That is, here we are back on the standard Hilbert space ${\cal H}_x = L^2(\mathbb{R},\rmd x)$, where expectation values of an observable are given in the usual way
\be
\langle \hat{A} \rangle_{\psi}  = \langle \psi| \hat{A}|  \psi \rangle =\int_{-\infty}^{\infty} \rmd x\, \psi^* (x)\bigl(\hat{A}\psi\bigr) (x)
\ee
However, the corresponding GUP Hamiltonian is not a standard Schr\"odinger Hamiltonian as it is of the form
\be\label{Hx}
H_x := -\frac{1}{2M\lambda^2} \tanh^2 \lb \lambda\p_x \rb  + V \lb x \rb \,.
\ee

\section{The GUP derivative and its properties}
The objective of this section is to present some properties of the GUP derivative which we define by
\be\label{Dxdefinition}
D_x:= \frac{1}{\lambda} \tanh \lb \lambda \p_x \rb= -\frac{\rmi}{\lambda}\tan(\rmi\lambda\p_x)=-\rmi\va^{-1}(\rmi\p_x)\,.
\ee
Here we note that the above operator is to be understood in terms of the power series
\be\label{tanhSeries2}
D_x  =\sum_{n=0}^{\infty}\frac{2^{2n+2}(2^{2n+2}-1)}{(2n+2)!} \,B_{2n}\,\lambda^{2n}\partial^{2n+1}_x
\ee
and thus is a linear operator. With this definition
the GUP Hamiltonian \eqref{Hx} now takes the simple form
\be\label{Hx2}
H_x = -\frac{1}{2M} D^2_x   + V \lb x \rb \,.
\ee
It is obvious that on ${\cal H}_x$ this derivative is anti-symmetric, i.e.\  $D_x^\dag = -D_x$. This implies that the corresponding momentum operator is symmetric, cf.\ \eqref{xReps},
\be
\pph = -\rmi D_x=\frac{1}{\rmi\lambda} \tanh \lb \lambda \p_x \rb\,,\qquad \pph^{\dag}=\pph
\ee
but it is not self-adjoint \cite{Pedram2012,Ref9}.  In fact, as we will see below, the eigenfunctions of $\hat{P}$ do not form a complete set in ${\cal H}_x$.

At a later stage we will also need to look at integrals of such derivatives, that is, integrals of the form
\be
\int_{a}^{b} \rmd x \, (D_x g)(x) = \sum_{n=0}^{\infty}\frac{2^{2n+2}(2^{2n+2}-1)}{(2n+2)!} \,B_{2n}\,\lambda^{2n}
\left[ g^{(2n)}(b) - g^{(2n)}(a)\right]\,.
\ee
Here $g^{(2n)}$ denotes the $2n$-th derivative of the function $g$. {\red In particular, in the limit  $a\to b$ the vanishing of above expression implies the continuity of $g$ (and all its even derivatives) at $b$.}


As an application, let us consider the stationary Sch\"odinger equation for a potential $V(x)$ being bounded but not necessarily continuous at say $x=x_0$ and integrate it over an interval containing $x_0$,
\be
-\frac{1}{2M}\int_{x_0 -\varepsilon}^{x_0 +\varepsilon} \rmd x\, D^2_x\psi(x) = \int_{x_0 -\varepsilon}^{x_0 + \varepsilon} \rmd x\, \Bigl[E-V(x)\Bigr]\psi(x)\,.
\ee
In the limit $\varepsilon\to 0$ the right-hand side vanishes and thus we arrive at a continuity condition for the even derivatives of the function $g=D_x\psi$. That is, the GUP derivative of the wave function at $x=x_0$, where the potential is bounded but not necessary continuous, should be continuous.
\be\label{GUPIntDx}
(D_x \psi)(x_0-0) = (D_x \psi)(x_0+0)\,.
\ee
We will utilise this condition when solving the problem where the potential characterises a box of finite depth.
\\

\noindent{\it GUP exponential:}\\
Here we begin by defining the GUP exponential function as follows
\be \label{GUPe}
e_{\beta} ( a; x) := \exp\left\{-\rmi \va (\rmi a) x \right\}\,,\qquad a\in \mathbb{C}\,,
\ee
{\red with $\va$ as previously defined in \eqref{varphi}.}
Using the well-know  relation $ f(\p_x )\,\rme^{ \rmi \alpha x } =  f\lb \rmi \alpha  \rb \rme^{  \rmi \alpha x }$, one immediately finds the relation
\be
 D_x e_{\beta} ( a; x)  = a e_{\beta} ( a; x)\,.
\ee
That is, the GUP exponential \eqref{GUPe} is an eigenfunction of the GUP derivative with eigenvalue given by its first parameter.
One can easily check the following relations for the GUP exponential;
\be
e_{\beta} ( a; 0)=1\,,\qquad  e_{\beta} ( 0; x)=1\,,\qquad e_{\beta} ( a; x)  e_{\beta} ( a; y)=  e_{\beta}( a ; x+y)\,.
 \ee
Hence, the GUP exponential obeys the usual functional relation of the standard exponential function in its second parameter.

As a first application of the GUP exponential, let us consider the GUP wave equation given by
\be\label{GUPwaveeq}
 ( D_x^2 + \omega^2 ) y (x) =0\,,\qquad \omega >0\,.
\ee
Obviously, the two linearly independent solutions are the  plane waves
\be
 e_{\beta} (\rmi\omega ; x)\qquad\mbox{and}\qquad  e_{\beta} (\rmi\omega ; x)\,.
\ee
The wavelength $\ell$ of these waves is given by $\ell = \frac{2\pi}{\va(\omega)}$ as $e_{\beta} (\rmi\omega ; x+\ell)= e_{\beta} (\rmi\omega ; x)$. Here we note that due to the boundedness of the deformation function $\va(\omega)<\frac{\pi}{2\lambda}$, the wave length is bounded from below by $\ell>4\lambda$. This is an obvious consequence of minimal length implied by GUP.

Finally we conclude this discussion by introducing the GUP cosine and GUP sine functions following the usual definitions
\be
C_{\beta} ( \omega; x) : = \frac{1}{2} \left[ e_{\beta} ( \rmi \omega; x) + e_{\beta} ( - \rmi \omega; x)\right]
 = \cos \left( \va (\omega )x \right)\,,
\ee
\be
S_{\beta} ( \omega; x) := \frac{1}{2\rmi} \left[ e_{\beta} ( \rmi \omega; x) - e_{\beta} ( - \rmi \omega; x)\right]
 = \sin \left( \va(\omega) x \right)\,,
\ee
which both are also solution of the GUP wave equation \eqref{GUPwaveeq}.

\section{Gaussian wave packets and GUP Fourier transformation}
{\it GUP momentum eigenfunctions:}\\
With the above exponential being an eigenfunction of the GUP derivative, it also serves as an eigenfunction for the momentum operator $\hat{P}=-\rmi D_x$. That is,
the GUP momentum eigenfunctions satisfying
\be
\pph u_{P}(x) = P u_{P}(x)\,,\qquad P\in\mathbb{R}\,,
\ee
are explicitly given by plane waves with a dispersion relation characterise by the deformation function $\va$,
\be\label{uP}
 u_{P}(x) = \frac{1}{ \sqrt{ 2 \pi ( 1 + \lambda^2 P^2)} } e_{\beta} ( i P; x)=\sqrt{\frac{\va'(P)}{2\pi}}\,\rme^{\rmi\va(P)x}\,.
\ee
Here we remark that the wave length $ \ell = \frac{2 \pi}{\va( P) }$ represents the GUP-corrected De Broglie relation.
Noting that $\va'(P)\delta(\va(P)-\va(P)) = \delta(P-P')$, one can easily check that they are properly normalised and othogonal
\be\label{Uorth}
\int_{-\infty}^{\infty} \rmd x\, u^*_{P}(x) u_{P'}(x) = \delta ( P - P')\,.
\ee
However, they do not form a complete set on ${\cal H}_x$ as
\be\label{isucomplete}
\int_{-\infty}^{\infty} \rmd P\, u^*_{P}(x) u_{P}(x') = \frac{1}{2\pi}
\int_{-\frac{\pi}{2\lambda}}^{\frac{\pi}{2\lambda}}\rmd \va\, \rme^{-\rmi\va(x-x')}= \frac{\sin\left(\frac{\pi}{2\lambda}(x-x')\right)}{\pi(x-x')}
\ee
and hence only in the limit $\lambda\to 0$ an exact completeness relation is achieved. This implies that a decomposition of any function into above momentum eigenfunctions cannot resolve details which are within a range of the width of above $\sin$-function. That is, details of order $\Delta x\sim \lambda$ become invisible in a momentum decomposition. See the appendix for a calculation of an approximate completeness relation up to first order in $\beta=\lambda^2$.\\

\noindent{\it Gaussian wave packet:}\\
Let us consider a normalised Gaussian wave function
\be
\phi_0(x) = \frac{1}{( 2 \pi \sigma^2)^{1/4} }e^{ -\frac{x^2}{4 \sigma^2}}
\ee
with a vanishing mean value $\langle x\rangle_{\phi_0}=0$ and variance
 \be
\Delta x = \sqrt{ \langle x^2 \rangle_{\phi_0} - \langle x\rangle_{\phi_0}^2 } =\sigma>0\,.
\ee
Despite the fact that the momentum eigenfunctions are not complete, we may analyse their components contained in the Gaussian wave function. The $P$-component is given by
\be\label{F}
g_0(P) :=   \int_{-\infty}^{\infty} \rmd x\, \phi_0(x)u^*_{P}(x) =
\int_{-\infty}^{\infty} \rmd x\, \phi_{\red 0}(x) \, \sqrt{\frac{\va'(P)}{2\pi}}\,\rme^{-\rmi\va(P)x}\,,
 \ee
which is trivially integrated and results in
\be
g_0(P) = \left(\frac{2}{\pi}\right)^{1/4}\, \sqrt{\sigma\va'(P)}\,\rme^{-[\sigma \va(P)]^2}= \left(\frac{2}{\pi}\right)^{1/4}\, \, \sqrt{\frac{\sigma}{1+\lambda^2 P^2}}\,\exp\left\{-\sigma^2 [\va(P)]^2\right\}\,.
\ee
As observed above, the momentum eigenfunctions are not complete. Therefore, we expect that the components $g_0$ do not contain the full information contained in $\phi_0$.
In order to estimate the loss when decomposing the Gaussian wave function into its $P$-components let us consider the quantity
\be
\begin{array}{rl}
\displaystyle
\int_{-\infty}^{\infty}\rmd P\, |g_0(P)|^2 &= \displaystyle \sqrt{\frac{2}{\pi}}\,\sigma \int_{-\infty}^{\infty}\rmd P\, \va'(P) \rme^{-2\sigma^2\va^2(P)}\\[4mm]
&= \displaystyle \frac{2}{\sqrt{\pi}} \int_{0}^{\frac{\pi\sigma}{\sqrt{2} \lambda}}\rmd t\,\rme^{-t^2} = {\rm erfc}\left(\frac{\pi\sigma}{\sqrt{2} \lambda}\right)\,.
\end{array}
\ee
where ${\rm erfc}$ denotes the complementary error function. If we assume that the width of our Gaussian is sufficiently large,\footnote{\red Physically this means that we are NOT exploring regions close to the Planck length. So this formalism allows us to describe "semiclassical" states in respect to full quantum gravity.
In other words, with the GUP we are closer to quantum gravity than with ordinary QFT, but not yet so close.} i.e.\ $\sigma \gg \lambda$, the asymptotic relation
\be\label{erfcasym}
{\rm erfc\,}(z)= \frac{\rme^{-z^2}}{z\sqrt{\pi}}\left(1+O(z^{-2})\right)
\ee
indicates that the loss due to the momentum decomposition becomes exponentially small
\be
\int_{-\infty}^{\infty}\rmd P\, |g_0(P)|^2 = 1-\sqrt{\frac{2}{\pi^3}}\,\frac{\lambda}{\sigma}\,\rme^{-\frac{\pi^2\sigma^2}{2\lambda^2}}\bigl(1+O(\lambda^2)\bigr)\,.
\ee
This implies that for a Gaussian with $\sigma \gg \lambda$, or more generally for sufficiently smooth wave functions, we may ignore the loss due to the incompleteness of the momentum eigenfunctions as the error will be exponentially small.\footnote{\red Recall that $p_0=\frac{\pi}{2\lambda}$ represents the cutoff of the canonical momentum as discussed in section 2. Hence this cutoff needs to obey $p_0\gg \frac{\pi}{2\sigma}$.} To confirm this, let us reconstruct the Gaussian from its $P$-components by
\be
\begin{array}{rl}
\tilde{\phi}_0(x)& := \displaystyle \int_{-\infty}^{\infty}\rmd P\, g_0(P)\,u_P(x) = \frac{1}{(2\pi)^{1/4}}\sqrt{\frac{\sigma}{\pi}}\int_{-\infty}^{\infty} \rmd P\, \va'(P)\,\rme^{\rmi\va(P)x}\rme^{-\sigma^2\va^2 (P)}\\
&= \displaystyle\frac{\rme^{-x^2/4\sigma^2}}{(2\pi\sigma^2)^{1/4}}\frac{\sigma}{\sqrt{\pi}}\int_{-\pi/2\lambda}^{\pi/2\lambda} \rmd \va\,\exp\left\{-\sigma^2 \left(\va +\rmi\frac{x}{2\sigma^2}\right)^2\right\}  \\
&= \displaystyle \phi_0(x) - \frac{1}{(2\pi\sigma^2)^{1/4}}\frac{2}{\sqrt{\pi}}\int_{\frac{\pi\sigma}{2\lambda}}^{\infty}\rmd t\,\rme^{-t^2}\cos( xt/\sigma)\,.
\end{array}
\ee
The last integral can be estimated as follows
\be
\left|\frac{2}{\sqrt{\pi}}\int_{\frac{\pi\sigma}{2\lambda}}^{\infty}\rmd t\,\rme^{-t^2}\cos( xt/\sigma)\right|\leq
\frac{2}{\sqrt{\pi}}\int_{\frac{\pi\sigma}{2\lambda}}^{\infty}\rmd t\,\rme^{-t^2}\left|\cos( xt/\sigma)\right|\leq {\rm erfc\,}\left(\frac{\pi\sigma}{2\lambda}\right)\,.
\ee
Again the result is an exponentially small correction to the original Gaussian under our assumption $\sigma\gg\lambda$. Hence, as we are only interested in corrections being of $O(\lambda^2)$, we may ignore those exponentially small errors.\\

\noindent{\it GUP Fourier transformation:}\\
Based on above discussion, we propose an approximate Fourier transformation for smooth wave functions $\phi$ as follows
\be
g(P) ={\cal F}( \phi(x)) := \int_{-\infty}^{\infty} \rmd x\, \phi(x) u^*_P(x) =\int_{-\infty}^{\infty} \rmd x\,\phi(x) \sqrt{\frac{\va'(P)}{2 \pi}} \rme^{-\rmi\va(P)x}
 \ee
with inverse transformation given by
\be
\tilde{\phi}(x) = {\cal F}^{-1}( g(p)):= \int_{-\infty}^{\infty} \rmd P\, g(P) u_P(x)=\int_{-\infty}^{\infty} \rmd P\,g(P) \sqrt{\frac{\va'(P)}{2 \pi}} \rme^{\rmi\va(P)x}\,.
\ee
Here for a smooth function $\phi$, which varies on large scales in the sense
\be
\langle x^2\rangle_\phi - \langle x\rangle_\phi^2 \gg \beta\,
\ee
we expect
\be
\tilde{\phi}(x) = \phi(x)+ O(\rme^{-1/\beta})\,.
\ee
In going forward we will identify both function, i.e., $\tilde{\phi}(x) = \phi(x)$. We will call the approximate Fourier transformation with $\tilde{\phi}(x)$ replaced by $\phi(x)$ the GUP Fourier transformation.
In that view, expectation values of any function say $A(P)$ are then approximately given by
\be
\langle A \rangle_{g_0} :=  \int_{-\infty}^{\infty}\rmd P\, A(P) |g_0(P)|^2\,.
\ee
Due to the symmetry of our Gaussian momentum distribution $g_0(P)$, the expectation value of $P$ as well as $\va(P)$ vanishes,
\be
\langle P\rangle_{g_0} = 0\,,\qquad \langle \va(P)\rangle_{g_0}=0\,,
\ee
but their variances $\Delta P = \sqrt{\langle P^2 \rangle_{g_0}}$ and $\Delta \va =\sqrt{ \langle \va^2 \rangle_{g_0}}$ are non-zero. In fact, we explicitly have
\be
\Delta \va =\frac{1}{2\sigma}
\ee
resulting in the uncertainty relation
\be
\Delta x \Delta \va=\frac{1}{2}\,.
\ee
However, the integral to calculate  $\langle P^2 \rangle_{g_0}$ cannot be evaluated  in closed form. Here we recall the relation \eqref{varphi}, which can be used to express $P$ in terms of $\va$. Namely, we have
\be
P=\frac{1}{\lambda}\,\tan(\lambda\va)\,,
\ee
which may be expressed in terms of a power series in $\va$ utilising the relation \eqref{tanhSeries2},
\be\label{Pseries}
P=\va\sum_{n=0}^{\infty}\frac{2^{2n+2}(2^{2n+2}-1)}{(2n+2)!} \,B_{2n}\,(-1)^n \beta^{n} \va^{2n}\,.
\ee
Hence $\langle P^2 \rangle_{g_0}$ can be expressed as a sum of higher-order Gaussian moments $\langle\va^{2m}\rangle_{g_0}= \frac{(2m-1)!!}{(2\sigma)^{2m}}$. To first order in $\beta$ we have
$P=\va+\frac{\beta}{3}\va^3+O(\beta^2)$ resulting in
\be\label{P2_0}
\langle P^2\rangle_{g_0}= \langle \va^2\rangle_{g_0} +\frac{2}{3}\beta \langle \va^4\rangle_{g_0} +O(\beta^2)= \frac{1}{4\sigma^2}+ \frac{\beta}{8\sigma^4}+O(\beta^2)\,.
\ee
This implies the uncertainty relation
\be\label{Uncertain_t=0}
\Delta P \Delta x = \frac{1}{2} \lb 1 + \frac{\beta}{ 4 (\Delta x)^2}+O(\beta^2) \rb = \frac{1}{2} \lb 1 + \beta \Delta P^2+O(\beta^2)\rb\,,
\ee
indicating that a Gaussian wave packet is to first order in $\beta$ a minimal uncertainty state according to \eqref{Uncertainty2}.

\section{Time evolution of a Gaussian wave function}
Now let us consider the time evolution of the free Gaussian wave packet. This is most suitable done via the Fourier transform, which in the absence of any external potential evolves in time
according to
\be
g_t(P)= \rme^{ - \frac{ \rmi P^2 t}{ 2 M}}g_0(P)\,.
\ee
Obviously, $|g_t(P)|^2=|g_0(P)|^2$ and hence $\left\langle P^2\right\rangle_{g_t}=\left\langle P^2\right\rangle_{g_0}$. That is, it is given by the initial expectation value \eqref{P2_0}.

By applying the inverse Fourier transformation we obtain the wave function at time $t>0$ as follows
\be\label{phit}
\begin{array}{rl}
\phi_t (x) & =\displaystyle \int_{-\infty}^{\infty} \rmd P\, g_t(P) u_P(x)\\
           & =\displaystyle  \left(\frac{\sigma^2}{2\pi^3}\right)^{1/4}\int_{-\infty}^{\infty} \rmd P\,\va'(P) \,\exp\left\{-\sigma^2\va^2(P)+\rmi\va(P)x- \frac{ \rmi P^2 t}{ 2 M}\right\}\,.
\end{array}
\ee
In general this integral may not be evaluated in closed form. However, using the power series for
$P=\frac{1}{\lambda}\tan(\lambda\varphi)= \varphi +\frac{1}{3}\lambda^2\varphi^3+O(\lambda^4)$
we may arrive at a power series in $\lambda^2$ with coefficients given by the moments of the Gaussian $\va$-distribution. Considering terms up to first order in $\lambda^2$ we have
\be
\rme^{-\rmi\frac{P^2}{2M}t}=
\rme^{-\rmi\sigma^2\varphi^2 \frac{t}{\tau}}\left( 1-\frac{\rmi}{3}\varphi^4\sigma^4\frac{\lambda^2}{\sigma^2}\frac{t}{\tau} +O\left(\frac{\lambda^4}{\sigma^4}\frac{t^2}{\tau^2}\right)\right)\,,
\ee
where we introduced the time scale $\tau:= 2M\sigma^2$. Now above integral \eqref{phit} is reduced to the evaluation of even momenta of the form
\begin{equation}\label{va2m}
  \langle\varphi^{2m}\rangle_{g_t}:= \sqrt{\frac{2\sigma^2}{\pi}}{\displaystyle \int_{-p_0}^{p_0}}\rmd \varphi\,\varphi^{2m}\rme^{-2\sigma^2\varphi^2} =
\frac{(2m-1)!!}{2^{2m}\sigma^{2m}}+\rme^{-\frac{\pi^2\sigma^2}{2\lambda^2}} O\left(\frac{\sigma^{2m-1}}{\lambda^{2m-1}}\right)\,.
\end{equation}
Again we have utilised the asymptotic form \eqref{erfcasym} of the complementary error function in the last step. The result reads
\begin{equation}\label{Phit1}
\phi_t(x) = \phi_t^{0}(x)\left[1-\rmi\frac{\lambda^2}{\sigma^2}\frac{t}{\tau}
  \left( \frac{1}{12}\frac{\sigma^8}{\sigma_t^8}\frac{x^4}{\sigma^4}- \frac{\sigma^6}{\sigma_t^6}\frac{x^2}{\sigma^2} + \frac{\sigma^4}{\sigma_t^4}\right) \right] +O\left(\frac{t^2}{\tau^2}\frac{\lambda^4}{\sigma^4}\right)\,,
\end{equation}
where $\phi^0_t (x)$ denotes the standard undeformed (i.e.\ $\lambda = 0$) Gaussian wave function evolving in time. In addition,   we have introduced the time-dependent width
\be
\sigma^2_t := \sigma^2 + \frac{\rmi t}{2M}=\sigma^2\left(1+\rmi\frac{t}{\tau}\right)\,,
\ee
characterising the complex width of the time-dependent undeformed Gaussian wave packet.

Let us consider a short time interval where $t\ll\tau$ and consider the width to first order in time, $\frac{\sigma}{\sigma_t}=1-\rmi \frac{t}{\tau}+O(\frac{t^2}{\tau^2})$. Then we observe
\begin{equation}\label{Phit2}
  |\phi_t(x)|^2 = |\phi_t^{0}(x)|^2 \left[ 1 - 2\frac{\lambda^2}{\sigma^2}\frac{t^2}{\tau^2}\left( \frac{2}{3}\frac{x^4}{\sigma^4} -6 \frac{x^2}{\sigma^2}+4\right)+O\left(\frac{t^3}{\tau^3}\frac{\lambda^2}{\sigma^2}\right) + O\left(\frac{t^2}{\tau^2}\frac{\lambda^4}{\sigma^4}\right)\right]
\end{equation}
and conclude with the well-known results $\langle x^2\rangle_{\phi^0_t}=\sigma^2(1+\frac{t^2}{\tau^2})$ and
$\langle x^4\rangle_{\phi^0_t}=3\sigma^4(1+\frac{t^2}{\tau^2})^2$ that $\phi_t(x)$ is up to the order indicated in \eqref{Phit2} already well normalized. Therefore we can calculated
\begin{equation}\label{x2t}
  \langle x^2 \rangle_{\Phi_t}=\int_{-\infty}^{\infty}\rmd x\,x^2|\Phi_t(x)|^2 =
  \sigma^2\left( 1+\frac{t^2}{\tau^2} +8\frac{\lambda^2}{\sigma^2}\frac{t^2}{\tau^2} +O\left(\frac{t^3}{\tau^3}\frac{\lambda^2}{\sigma^2}\right) + O\left(\frac{t^2}{\tau^2}\frac{\lambda^4}{\sigma^4}\right)\right)
\end{equation}
and conclude the time-dependent uncertainty relation
\be
(\Delta x)^2_{\phi_t} (\Delta P)^2_{\phi_t} = \frac{1}{4}\left[1+\frac{t^2}{\tau^2}+\frac{\lambda^2}{\sigma^2}+ \frac{17}{2}\frac{\lambda^2}{\sigma^2} \frac{t^2}{\tau^2}+O\left(\frac{t^3}{\tau^3}\frac{\lambda^2}{\sigma^2}\right) + O\left(\frac{t^2}{\tau^2}\frac{\lambda^4}{\sigma^4}\right) \right]\,,
\ee
which coincides for $t=0$ with the result \eqref{Uncertain_t=0}.

\section{Examples}
In this section we will discuss a few special cases of a GUP particle interacting with an external scalar potential $V(x)$.
These are the one-dimensional box and an attractive potential well with finite depth.\\

\noindent{\it The one-dimensional box problem:}\\
Here the potential considered represents a box of linear extension $L$ with infinite walls at its boundaries,
\be
V (x) :=\left\{
\begin{array}{ll}
  0 &  0<x< L \\
  \infty ~~& \mbox{elsewhere}
\end{array}\right. \,.
\ee
The corresponding stationary Schr\"odinger equation then reads
\be
-\frac{D_x^2}{2M} \psi(x) = E \psi(x)\,,
\ee
where we assume the Dirichlet boundary conditions $\psi(0)=0=\psi(L)$. Setting $E=\kappa^2/2M$ with $\kappa>0$, this may be rewritten in the form
\be
\left( D_x^2 + \kappa^2\right)\psi (x) = 0\,,
\ee
which is the wave equation discussed at the end of section 3. Therefore we may write the general solution  in terms of the GUP sine and cosine functions,
\be
\psi (x) =  c_1 C_{\beta} ( \kappa ; x) +  c_2 S_{\beta}( \kappa ; x)\,.
\ee
The boundary condition at $x=0$ requires $c_1 = 0$ and $c_2= \sqrt{\frac{2}{ L} }$ results in a proper normalisation. The boundary condition at $x=L$ then requires
$\sin(\va(\kappa)L)=0$, which provides us with the quantisation condition for $\kappa$,
\be
L\va(\kappa) = n\pi\,.
\ee
Here we recall that $\va$ is the principle branch of the arctan and hence $|\va(\kappa)|<\frac{\pi}{2\lambda}$. This results in a finite number of bound states.
\be
\kappa_n = \frac{1}{\lambda} \tan \lb \frac {\lambda}{L} n \pi \rb\,,\qquad n=1,2,3,\ldots, n_{\rm max}<\frac{L}{2\lambda} \,.
\ee
The corresponding wave functions read
\be
\psi_n (x) = \sqrt{\frac{2}{ L} } \sin \frac{ n \pi}{L} x
\ee
and are identical in form to the solutions of the same problem in standard quantum mechanics when $\lambda = 0$. The difference is visible only in the spectrum given by the eigenvalues
\be\label{EnBox}
 E_n = \frac{1}{ 2M {\lambda^2}} \tan^2  \lb \frac {\lambda}{L} n \pi \rb\,,\qquad n=1,2,3,\ldots, n_{\rm max}<\frac{L}{2\lambda} \,.
\ee
That is, for a width $L<2\lambda$ there are no solutions and for any finite $L$  we have a finite number of bound states. Note that for $n\to \frac{L}{2\lambda}$ we have $E_n\to\infty$, which in essence tells us that we need an infinite amount of energy to squeeze a GUP particle into a box of the size of $2\lambda$ or less.

The expectation values of position and momentum as well as their square expectation values for an eigenstate $\psi_n$ are given by
\be
\langle  x \rangle= \frac{L}{2}\,,\quad \langle  P \rangle =  0\,,\quad
\langle  x^2 \rangle= L^2 \lb \frac{1}{3} - \frac{1}{ 2 n^2 \pi^2} \rb\,,\quad\langle  P^2 \rangle=  \frac{1}{\lambda^2} \tan^2 \lb \frac {\lambda}{L} n \pi \rb\,.
\ee
Thus, the uncertainty relation in state $\psi_n$ reads
\be
\Delta x \Delta P = \frac{L}{\lambda} \tan \lb \frac {\lambda}{L} n \pi \rb    \sqrt{ \frac{1}{12} - \frac{1}{2 (n \pi)^2 } }\,.
\ee
Finally let us mention that above solution for a small $\beta$ agrees with the findings of ref.\  \cite{Nozari2006,Pedram2012}, which both use the approximated derivative \eqref{Dxapprox}. However, due to this approximated GUP derivative the first paper misses the essential point that the number of eigenvalues is finite as indicated in \eqref{EnBox}. In \cite{Pedram2012} a semiclassical analysis resulted in an estimate of the number of bound states given by $ n_{\rm max}<\frac{2L}{3\lambda}$ which is slightly higher then our exact result \eqref{EnBox}. \\

\noindent{\it The attractive potential well:}\\
Now we consider a one-dimension potential well with a finite depth characterised by a parameter $V_0<0$ and a width given by $2a>0$.
The potential thus reads
\be
V (x) :=\left\{
\begin{array}{cl}
  0 &  |x| > a \\
  ~~ V_0 ~~& |x| < a
\end{array}\right. , \qquad a>0\,, \qquad V_0 <0\,,
\ee
and the time independent Schr\"odinger equation becomes
\be
- \frac{1}{2M} D_x^2 \phi(x) = \bigl( E -V(x)\bigr)\phi(x)\,.
\ee
Here we only consider bound states, i.e. $V_0< E<0$, and introduce the two quantities $\kappa>0$ and $q>0$ given by
\be\label{aq}
E=-\frac{\kappa^2}{2M}<0 \,,\qquad E - V_0 = \frac{q^2}{2M} >0\,.
\ee
As the potential is symmetric in $x$, the corresponding eigenfunctions are either symmetric or anti-symmetric, $\phi^\pm(-x)=\pm\phi^\pm(x)$. They are then given by
\be
\phi^+(x)= \left\{
\begin{array}{ll}
  A \rme^{\rmi \va(\rmi\kappa)\,|x|} &  |x| > a \\
  B\cos(\va(q)x) & |x| < a
\end{array}\right.
\ee
and
\be
\phi^-(x)= \left\{
\begin{array}{ll}
  A\frac{x}{|x|}\rme^{\rmi \va(\rmi\kappa)\,|x|} &  |x| > a \\
  C\sin(\va(q)x) & |x| < a
\end{array}\right.\,,
\ee
respectively. The continuity of $\phi$ and $D_x\phi$ at $x=a$ then brings us to the two conditions
\be\label{kappa1}
\kappa=q\tan(\va(q)a)\qquad \mbox{and}\qquad \kappa=-q\cot(\va(q)a)
\ee
for the even and odd solutions, respectively. Both conditions are not explicitly solvable.
Let us introduce the dimensionless quantity  $\gamma:=a\sqrt{q^2 +\kappa^2}=\sqrt{2Ma^2|V_0|}$, which in essence is a measure for the strength of the potential. Then with $y:=a q$ above conditions become
\be\label{cond1}
\frac{\sqrt{\gamma^2 -y^2}}{y}=
\left\{
\begin{array}{ll}
  \tan\left(\frac{a}{\lambda} \arctan(\frac{\lambda}{a}y)\right)\qquad & \mbox{for}~\phi_+ \\[2mm]
  -\cot\left(\frac{a}{\lambda}\arctan(\frac{\lambda}{a}y)\right) & \mbox{for}~\phi_-
\end{array}
\right.\,.
\ee
\begin{figure}
  \centering
  \includegraphics[bb = 00 00 250 150, scale=0.75]{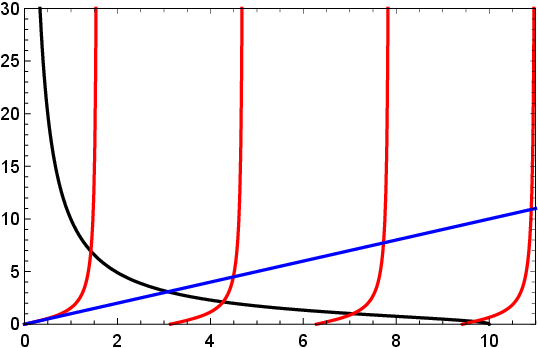}\qquad\qquad
  \includegraphics[bb = 00 00 250 150, scale=0.75]{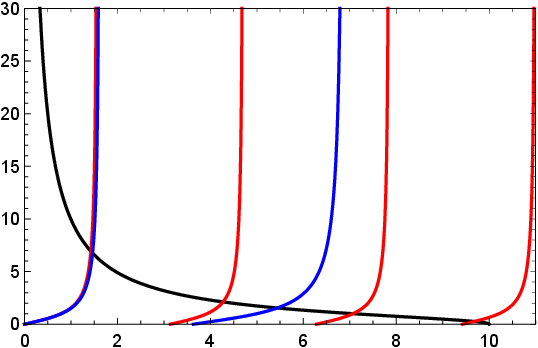}\\[10mm]
  \includegraphics[bb = 00 00 250 150, scale=0.75]{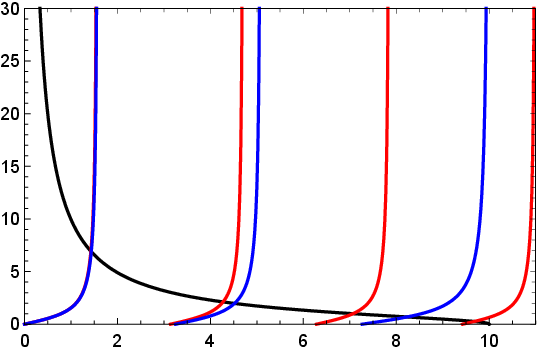}\qquad\qquad
  \includegraphics[bb = 00 00 250 150, scale=0.75]{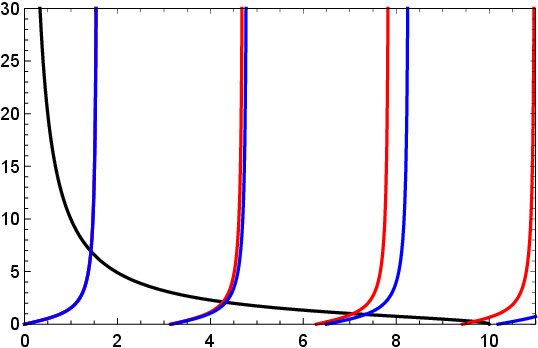}
  \caption{The graphical presentation of the even solutions of \eqref{cond1} for fixed $\gamma =10$ and $a/\lambda$ varying from $1,5,10,20$ from left upper corner till right lower one. The intersections of the black line with blue lines indicate the even solutions to the eigenvalue problem. The red lines indicate the undeformed case with $\lambda = 0$.}\label{Fig1}
\end{figure}
\begin{figure}
  \centering
  \includegraphics[bb = 00 00 250 150, scale=0.75]{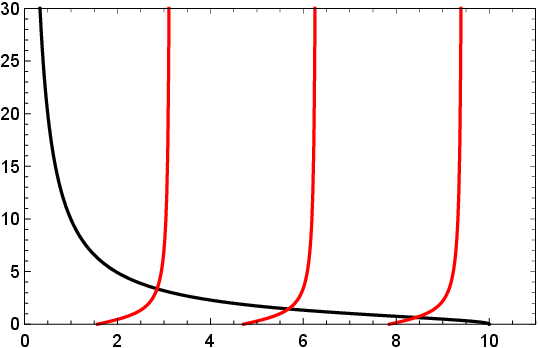}\qquad\qquad
  \includegraphics[bb = 00 00 250 150, scale=0.75]{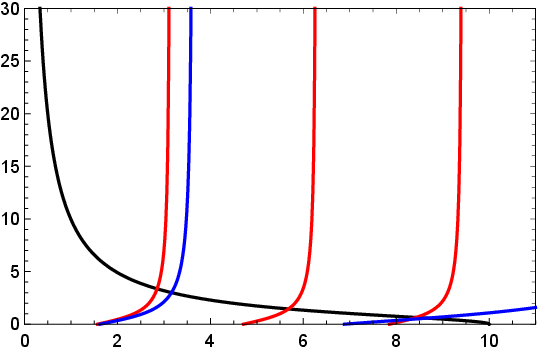}\\[10mm]
  \includegraphics[bb = 00 00 250 150, scale=0.75]{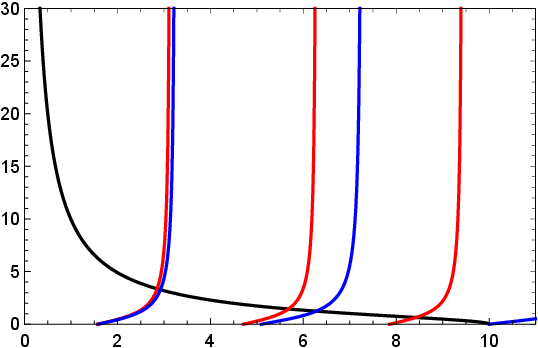}\qquad\qquad
  \includegraphics[bb = 00 00 250 150, scale=0.75]{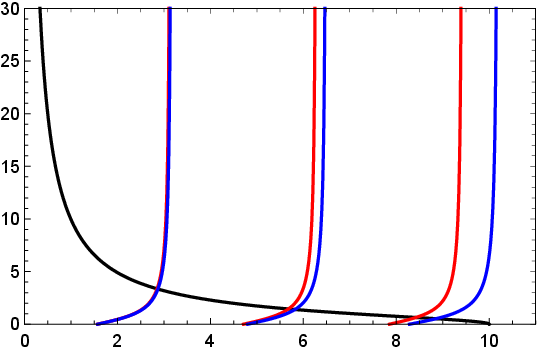}
  \caption{Same as figure \ref{Fig1} but for the odd solutions of \eqref{cond1} with same set of parameters. The case $a=\lambda$ in the left upper graph does not exhibit an odd eigenstate as discussed in the text.}\label{Fig2}
\end{figure}
It appears that there always exists a ground state. No excited states occur when $\gamma <\frac{a}{\lambda}\tan(\frac{\lambda}{a}\pi)$.
In figure 1 and 2 we present graphical solutions for both conditions \eqref{cond1} for fixed value $\gamma =10$ and $a/\lambda=1,5,10,20$, respectively. Figure 1 shows the even solutions related to the first condition in \eqref{cond1}. The black line presents the left-hand-side of this condition, whereas the blue lines show the right-hand-side of \eqref{cond1}. In addition, we added red lines indicating the undeformed case, see the leading term below in eq.\ \eqref{cond2}. The parameter $a/\lambda$ varies from $a/\lambda=1$ in the left upper graph to $a/\lambda=20$ in the right lower one. Figure 2 shows the same graphs but for the odd solutions in \eqref{cond1}. Both figures clearly show that with increasing width the GUP potential well accumulates more and more bound states, indicated by the intersections of the black and blue lines. For increasing $a$ the difference to the undeformed potential well becomes less invisible, in particular for the lower eigenvalues. For $a=20\lambda$ the ground state energy is essential identical to that of the undeformed case as the red curve completely disappeared  underneath the blue one.

For $a\gg \lambda$ one obtains the approximate conditions
\be\label{cond2}
\frac{\sqrt{\gamma^2 -y^2}}{y}=
\left\{
\begin{array}{ll}
  \tan y -\frac{1}{3}\frac{\lambda^2}{a^2}y^3 \left(1+\tan^2 y\right)+O(\frac{\lambda^4}{a^4})\qquad & \mbox{for}~\phi_+ \\[2mm]
  -\cot y -\frac{1}{3}\frac{\lambda^2}{a^2}y^3 \left(1+\cot^2 y\right)+O(\frac{\lambda^4}{a^4})  & \mbox{for}~\phi_-
\end{array}
\right.\,.
\ee
Clearly $\lambda = 0$ reproduces the standard textbook result indicated by the red curves in both figures.

For the special case $a=\lambda$ the above conditions \eqref{kappa1} allow for an analytic solution as they are reduced to
\be\label{kappa2}
\kappa=\lambda q^2 \qquad \mbox{and}\qquad \kappa=-\frac{1}{\lambda}\,.
\ee
As $\kappa>0$, we have to discard the second solution belong to an odd eigenstate and we remain with a single bound state with energy eigenvalue given by
\be
E_1=V_0+ \frac{1}{4M\lambda^2}\left(\sqrt{1-8V_0M\lambda^2} - 1\right)\,.
\ee
Here we observe that for $a=\lambda\to 0$ this approaches the value $E_1\to 0$ and the ground state disappears.

To conclude this discussion on the square well,  let us look at the limit of small $a$ and large $V_0$ such that the product $\alpha:=a|V_0|$ remains constant. That is, in the limit $a\to 0$ the box simulates a $\delta$ potential. For small $a$ relation \eqref{aq} results in a large $q$ approximately given by $q^2\approx\frac{2M\alpha}{a}$. For large $q$ a little exercise shows that $\tan(\va(q)a)\approx\frac{\pi}{2} \frac{a}{\lambda}$. Hence, via eq.\  \eqref{kappa1} we arrive at the ground state energy
$$
E_1\approx -\frac{a\alpha\pi^2}{4\lambda^2}\,.
$$
That is, in the limit $a\to 0$, which simulates a potential $V(x)\to 2\alpha\delta(x)$, the ground state disappears again. In other words, the Dirac $\delta$-potential does not have a bound state in gravitational quantum mechanics.

\section{Summary}
In this paper we have presented an algebraic approach to gravitational quantum mechanics being based on a deformed
Heisenberg algebra resulting in a generalized uncertainty principle. The GUP introduces a minimal length $\lambda$ being directly related to the deformation parameter $\beta=\lambda^2$, and has severe effects on the associated quantum models. For the investigation of such deformed quantum mechanical models, we studied some properties of the GUP derivative introduced in section 3. Our findings are essential ingredients to obtain exact solutions for several models in gravitational quantum mechanics.

The plane-wave solutions{\red were} shown to be characterised by the so-called GUP exponential function which allowed a superposition to form a Gaussian wave packet. It appears that details of such wave packets are lost when decomposing them into their plane wave components. However, such details are shown to be exponentially small and hence, can be ignored when looking into effects being of the order of $\beta$. This has let us to introduce the GUP Fourier transformation, which is an approximation of the standard Fourier transformation. This enabled us to look into the time evolution of a Gaussian wave packet, which was analysed to first order in $\beta=\lambda^2$. The corresponding uncertainty relation has been present explicitly to second order in time.

We further investigated several models where the GUP particle interacts with an external potential, for which we could solve exactly the corresponding eigenvalue problem with the help of the GUP exponential. For the particle enclosed in a box it turned out that the corresponding energy eigenfunctions are identical in form with those of the undeformed case. The effect of a minimal length only shows up in the corresponding spectrum which consists of a finite number of eigenvalues only. This is clearly due to the effect of a minimal length as for a box of size $L<2\lambda$ no energy eigenstates exist at all as they would require an infinite amount of energy.

The situation is different when considering a potential well with a finite depth $-V_0$ and a width of size $2a$.
Here again only a finite number of bound states exits but for any $a>0$ there at least is one bound state.
For large $a\gg \lambda$ the low-lying eigenvalues are very close to those of the undeformed case. For the special case $a=\lambda$ only one bound state exists whose eigenvalue could be given in closed form. The particular case where $a\to 0$ and $V_0\to -\infty$ such that $a|V_0|$ remains constant the well simulates a Dirac delta potential but does not exhibit  any bound state in that limit.

\section*{Appendix: Approximate completeness of GUP momentum eigenfunctions}
Let us reconsider the relation \eqref{isucomplete} which reads
\be\label{A1}
\int_{-\infty}^{\infty} \rmd P\, u^*_{P}(x) u_{P}(x') = \int_{-\infty}^{\infty} \rmd P\,\frac{\rme^{\rmi \va(P)(x-x')}}{2\pi(1+\beta P^2)}\,
\ee
Noting that $\va(P)= P - (\beta/3)P^3+O(\beta^2)$ we realise that the $P^3$-term in the exponent contributes only with a leading order in $\beta^2$ to \eqref{A1} due to its anti-symmetry. Hence the  leading term in $\beta$ comes from the measure $(1+\beta P^2)^{-1}=1- \beta P^2 +O(\beta^2)$. This brings us to the approximate completeness relation
\be
\begin{array}{rl}
\displaystyle
\int_{-\infty}^{\infty} \rmd P\, u^*_{P}(x) u_{P}(x') &\displaystyle= \frac{1}{2\pi} \int_{-\infty}^{\infty} \rmd P\,\rme^{\rmi P(x-x')}\left( 1- \beta P^2 +O(\beta^2)\right)\\[4mm]
&\displaystyle=\delta ( x - x') +\beta \delta'' ( x - x') +O(\beta^2).
\end{array}
\ee
That is, the first order correction term in $\beta$ is represented by a second-order derivative of the delta function.

\section*{Acknowledgements}
{\red The research of H.~H. was supported by the Q-CAYLE project, funded by the European Union-Next Generation UE/MICIU/Plan de Recuperacion, Transformacion y Resiliencia/Junta de Castilla y Leon (PRTRC17.11), and also by project PID2023-148409NB-I00, funded by MICIU/AEI/10.13039/501100011033. Financial support of the Department of Education of the Junta de Castilla y Leon and FEDER Funds is also gratefully acknowledged (Reference: CLU-2023-1-05).}
H.~H.\ would also like to thank the Department  of   Physics,   University   of   Hradec   Kr\'{a}lov\'{e} for their hospitality.

\section*{Declarations}

{\red The authors did not receive any other support from any organization for the submitted work besides that mentioned in above acknowledgment.}\\
The authors have no conflicts of interest to declare that are relevant to the content of this article.\\
Data Availability Statement: No Data associated in the manuscript.

\end{document}